\documentstyle[12pt,epsf]{article}

\newcommand{\nc}{\newcommand}
\nc{\fdiag}{0}
\nc{\bg}{B. Grz${{{\rm a}_{}}_{}}_{\hskip -0.18cm\varsigma}$dkowski}
\nc{\lsp}{\;\;\;\;\;\;\;\;}
\nc{\beq}{\begin{equation}}   \nc{\eeq}{\end{equation}}
\nc{\bea}{\begin{eqnarray}}   \nc{\eea}{\end{eqnarray}}
\nc{\baa}{\begin{array}}      \nc{\eaa}{\end{array}}
\nc{\bit}{\begin{itemize}}    \nc{\eit}{\end{itemize}}
\nc{\ben}{\begin{enumerate}}  \nc{\een}{\end{enumerate}}
\nc{\bce}{\begin{center}}     \nc{\ece}{\end{center}}
\nc{\non}{\nonumber}
\nc{\lumun}{\;{\hbox {pb}^{-1}}{\hbox {yr}^{-1}}}
\nc{\hc}{\hbox {h.c.}}
\nc{\re}{\hbox {Re}}
\nc{\im}{\hbox {Im}}
\nc{\etal}{\hbox{et al.}}
\def\MPL #1 #2 #3 {{\sl Mod.~Phys.~Lett.}~{\bf#1} (#3) #2}
\def\NPB #1 #2 #3 {{\sl Nucl.~Phys.}~{\bf B#1} (#3) #2}
\def\PLB #1 #2 #3 {{\sl Phys.~Lett.}~{\bf B#1} (#3) #2}
\def\PR #1 #2 #3 {{\sl Phys.~Rep.}~{\bf#1} (#3) #2}
\def\PRD #1 #2 #3 {{\sl Phys.~Rev.}~{\bf D#1} (#3) #2}
\def\PRL #1 #2 #3 {{\sl Phys.~Rev.~Lett.}~{\bf#1} (#3) #2}
\def\RMP #1 #2 #3 {{\sl Rev.~Mod.~Phys.}~{\bf#1} (#3) #2}
\def\ZPC #1 #2 #3 {{\sl Z.~Phys.}~{\bf C#1} (#3) #2}
\def\IJMP #1 #2 #3 {{\sl Int.~J.~Mod.~Phys.}~{\bf#1} (#3) #2}

\nc{\ra} {\rightarrow}
\nc{\cw}{\cos\theta_W}        \nc{\sw}{\sin\theta_W}
\nc{\ttbar}{t\bar{t}}
\nc{\bbbar}{b\bar{b}}
\nc{\tanb} {\tan \beta}
\nc{\twbdec} {t\rightarrow W^+ b}
\nc{\tbwbdec} {\bar{t} \rightarrow W^- \bar{b}}
\nc{\hprod} {e^+e^- \ra Z^\ast \ra H Z}
\nc{\epem} {e^+e^-}
\nc{\wpwm} {W^+W^-}
\nc{\tbar} {\bar{t}}
\nc{\bbar} {\bar{b}}
\nc{\wpp} {W^+}
\nc{\mt}{m_t}
\nc{\mts}{m_t^2}
\nc{\mw} {m_W}
\nc{\mws} {m_W^2}
\nc{\mz} {m_Z}
\nc{\mzs} {m_Z^2}
\nc{\mh} {m_H}
\nc{\mhs} {m_H^2}
\nc{\ma} {m_A}
\nc{\mas} {m_A^2}
\nc{\hdec}{H \ra t\bar{t}}
\nc{\ttbardec}{\ttbar \ra W^+W^-\bbbar}
\nc{\po}{\Phi_1}
\nc{\pod}{\Phi_1^\dagger}
\nc{\pht}{\Phi_2}
\nc{\phtd}{\Phi_2^\dagger}
\nc{\phtt}{{\tilde{\Phi}}_2}
\nc{\popo}{\po^\dagger\po}
\nc{\phtpt}{\pht^\dagger\pht}
\nc{\popt}{\po^\dagger\pht}
\nc{\phtpo}{\pht^\dagger\po}
\nc{\sq}{\sqrt{2}}
\nc{\nsd} {N_{SD}}
\nc{\ntt} {N_{tt}}

\def\gtino{\wt g_{3/2}}
\def\mgtino{m_{3/2}}
\def\mplanck{\mpl}
\def\mpl{M_{\rm P}}

\def\Eq#1{Eq.~(\ref{#1})}

\def\nmess{N_m}

\def\sq{\wt q}

\def\msq{m_{\sq}}

\def\slepl{\wt \ell_L}
\def\mslepl{m_{\slepl}}
\def\slepr{\wt \ell_R}
\def\mslepr{m_{\slepr}}

\def\sell{\wt e_L}
\def\msell{m_{\sell}}
\def\selr{\wt e_R}
\def\mselr{m_{\selr}}

\def\susy{{\rm SUSY}}

\def\cnone{\wt\chi^0_1}

\def\etmiss{/ \hskip-8pt E_T}
\def\etmin{/ \hskip-8pt E_T^{\rm min}}
\def\etjet{E_T^{\rm jet}}

\def\gl{\wt g}
\def\mgl{m_{\gl}}

\def\stopone{\wt t_1}

\def\mstopone{m_{\stopone}}

\def\sq{\wt q}

\def\msq{m_{\sq}}

\def\slepl{\wt \ell_L}
\def\mslepl{m_{\slepl}}
\def\slepr{\wt \ell_R}
\def\mslepr{m_{\slepr}}
\def\fbi{~{\rm fb}^{-1}}
\def\fb{~{\rm fb}}
\def\pbi{~{\rm pb}^{-1}}
\def\pb{~{\rm pb}}

\def\gev{~{\rm GeV}}
\def\tev{~{\rm TeV}}

\def\mt{m_t}

\def\mcnone{m_{\cnone}}

\def\wt{\widetilde}
\def\cpone{\wt \chi^+_1}

\def\mcpone{m_{\cpone}}

\def\stauone{\wt \tau_1}

\def\wtil{\widetilde}

\def\vev#1{\langle {#1}\rangle}

\def\lsim{\mathrel{\raise.3ex\hbox{$<$\kern-.75em\lower1ex\hbox{$\sim$}}}}
\def\gsim{\mathrel{\raise.3ex\hbox{$>$\kern-.75em\lower1ex\hbox{$\sim$}}}}

\def\pbi{~{\rm pb}^{-1}}
\def\fbi{~{\rm fb}^{-1}}
\def\fb{~{\rm fb}}
\def\pb{~{\rm pb}}

\def\gev{\,{\rm GeV}}
\def\tev{\,{\rm TeV}}
\def\wh{\widehat}
\def\wt{\widetilde}

\def\gl{\wt g}

\def\call{{\cal L}}

\def\etc{{\it etc.}}

\def\ifmath#1{\relax\ifmmode #1\else $#1$\fi}
\def\half{\ifmath{{\textstyle{1 \over 2}}}}
\def\threehalf{\ifmath{{\textstyle{3 \over 2}}}}

\def\etc{{\em etc.}}

\def\rts{\sqrt s}

\def\h{h}
\def\a{a}
\def\mh{m_{\h}}
\def\ma{m_{\a}}

\def\eg{{\it e.g.}}
\def\etal{{\it et al.}}
\def\epem{e^+e^-}

\def\lsim{\mathrel{\raise.3ex\hbox{$<$\kern-.75em\lower1ex\hbox{$\sim$}}}}
\def\gsim{\mathrel{\raise.3ex\hbox{$>$\kern-.75em\lower1ex\hbox{$\sim$}}}}
\def\@versim#1#2{\vcenter{\offinterlineskip
        \ialign{$\m@th#1\hfil##\hfil$\crcr#2\crcr\sim\crcr } }}

\def\ie{{\it i.e.}}

\def\gam{\gamma}

\def\nsd{N_{SD}}

\def\pbi{~{\rm pb}^{-1}}
\def\fbi{~{\rm fb}^{-1}}
\def\fb{~{\rm fb}}
\def\pb{~{\rm pb}}

\def\gev{\,{\rm GeV}}
\def\tev{\,{\rm TeV}}
\def\wh{\widehat}
\def\wt{\widetilde}

\def\gl{\wt g}
\def\mgl{m_{\gl}}

\def\stopone{\wt t_1}

\def\mstopone{m_{\stopone}}
\def\sq{\wt q}

\def\msq{m_{\sq}}

\def\slepl{\wt \ell_L}
\def\mslepl{m_{\slepl}}
\def\slepr{\wt \ell_R}
\def\mslepr{m_{\slepr}}

\def\hl{h^0}

\def\tanb{\tan\beta}

\def\mt{m_t}

\def\mz{m_Z}
\def\mw{m_W}
\def\mgut{M_U}

\def\cnone{\wt\chi^0_1}

\def\mcnone{m_{\cnone}}

\def\h{h}
\def\mh{m_{\h}}
\def\cpone{\wt \chi^+_1}

\def\mcpone{m_{\cpone}}

\def\stauone{\wt \tau_1}

\begin{document}
%
\font\fortssbx=cmssbx10 scaled \magstep2
\medskip
$\vcenter{
\hbox{\fortssbx University of California - Davis}
}$
\hfill
$\vcenter{
\hbox{\bf UCD-98-3} 
\hbox{January, 1998}
}$
\vspace*{1cm}
\begin{center}
{\large{\bf Probing Gauge-Mediated Supersymmetry Breaking Models at the Tevatron
       via Delayed Decays of the Lightest Neutralino}}\\
\rm
\vspace*{1cm}
{\bf  C.-H. Chen and J.F. Gunion}\\
\vspace*{1cm}
{\em Department of Physics, 
University of California, Davis, CA, USA }
\end{center}
\begin{abstract}
We quantitatively explore, in the context of the D0 detector
at the Tevatron, three very different techniques for
observing delayed decays of the
lightest neutralino of a simple gauge-mediated supersymmetry
breaking (GMSB) model to photon plus gravitino. It is demonstrated
that the delayed-decay signals considered can greatly increase the
region of general GMSB parameter space for which supersymmetry can
be detected. In the simple class of models considered, 
the combination of standard supersymmetry signals and delayed-decay
signals potentially yields at least one viable signal for nearly
all of the theoretically favored parameter space.
The importance, for delayed-decay signal detection,
of particular detector features and
of building a simple photon detector
on the roof of the D0 detector building is studied.
\end{abstract}
\vspace{5mm}

\setcounter{page}{0}
\thispagestyle{empty}
\section{Introduction}

Even in the context of the Minimal Supersymmetric Model (MSSM),
it has become increasingly apparent that there are many different
ways in which supersymmetry breaking can occur, leading to many different
possible phenomenologies. In particular, 
gauge-mediated supersymmetry breaking (GMSB) models have
attracted a great deal of interest in the last few years.
The theory of gauge-mediated supersymmetry breaking posits that
supersymmetry breaking is transmitted to the supersymmetric partners
of the Standard Model (SM) particles via the
SU(3)$\times$SU(2)$\times$U(1) gauge forces.
Two GMSB model-building approaches have been explored in the
literature.  
\bit
\item In hidden-sector models, the GMSB model consists of
three distinct sectors: a hidden sector
where supersymmetry is broken, a ``messenger'' sector containing
a singlet field and
messenger fields with SU(3)$\times$SU(2)$\times$U(1) quantum
numbers, and a sector containing
the fields of the MSSM \cite{basicgmsb}.  The
coupling of the messengers to the hidden sector generates 
supersymmetry-breaking in the messenger sector.
\item
In models of ``direct gauge mediation'' (sometimes called
direct-transmission models) \cite{dgm}, the GMSB model consists only of
two distinct sectors: a sector which is not only responsible for
supersymmetry breaking but also contains the messenger
fields; and the sector of MSSM fields.  
\eit
In both classes of models,
supersymmetry-breaking is transmitted to the MSSM sector via the SM
SU(3)$\times$SU(2)$\times$U(1)
gauge interactions between messenger fields
and the MSSM fields.  In particular, soft-supersymmetry-breaking masses for
the gauginos and squared-masses for the squarks and sleptons arise,
respectively, from one-loop and
two-loop diagrams involving the virtual exchange of messenger fields.

Let us consider further the simple hidden-sector models of the first class 
in which the communication of supersymmetry breaking in the hidden sector
to the messenger sector occurs via two-loop interactions involving a new gauge
group with coupling $g_m$ and in which the messenger sector consists of 
an SU(5) singlet superfield $\wh S$ and a certain effective number $\nmess$ of
complete SU(5) representations. In such models the boundary conditions
for the soft-supersymmetry-breaking masses are set at the mass scale
of the messenger sector $M_m$ and take a particularly simple form.
For the gaugino Majorana mass terms one finds~\cite{basicgmsb}:
\beq 
M_i(M_m)=k_i\nmess g\left({\Lambda\over M_m}\right) 
{\alpha_i(M_m)\over 4\pi}\Lambda\,,
\label{gmsbmi}
\eeq
where $k_2=k_3=1,k_1=5/3$ and
$\nmess\leq 4$ is required to avoid Landau poles.
$\Lambda=\vev{F_S}/\vev{S}$ is the ratio of the vacuum expectation values
of the auxiliary and scalar field components of the chiral field $\wh S$.
Note that the gaugino masses have the same relative values
as if they were unified at the grand-unification scale $\mgut$
even though the actual initial conditions are set at scale $M_m$.
In addition,
scalar masses are expected to be flavor independent since the mediating
gauge forces are flavor-blind.  In the simple models we consider here,
they take the form:
\beq
m_i^2(M_m)=2\Lambda^2\nmess f\left({\Lambda\over M_m}\right)
\sum_{i=1}^3 c_i \left({\alpha_i(M_m)\over 4\pi}\right)^2
\label{gmsbmsq}
\eeq
with $c_3=4/3$ (for color triplets), $c_2=3/4$ (for weak doublets) and 
$c_1={5\over 3}\left({Y\over 2}\right)^2$ (in 
the normalization where $Y/2=Q-T_3$). 
To avoid negative mass-squared for bosonic members of the messenger sector 
$M_m/\Lambda> 1$ is required; $M_m/\Lambda\geq 1.1$ is
preferred to avoid fine-tuning, for which $f(\Lambda/M_m)\simeq 1$ and
$1\leq g(\Lambda/M_m)\leq 1.23$. For $M_m/\Lambda\geq 2$, 
$1\leq g(\Lambda/M_m)\leq
1.045$. Here, we will consider masses obtained with $g=f=1$.
The results above at scale $M_m$ must be evolved down to the scale of the
actual sparticle masses, denoted $Q$.
The resulting gaugino masses are given
by replacing $\alpha_i(M_m)$ in Eq.~(\ref{gmsbmi}) by $\alpha_i(Q)$.
Evolution of the sfermion masses is detailed in Ref.~\cite{martin}.
Most important is the ratio
$\mslepr/M_1=\sqrt{6/5}\sqrt{r_1/\nmess-(5/33)(1-r_1)}$,
where $r_1=(\alpha_1(M_m)/\alpha_1(Q))^2$. For the very broad
range of $Q\geq\mz$
and $M_m\leq 3\times 10^6\tev$, $1<r_1\lsim 1.5$, in which case
the lightest of the sparticle partners of the SM
particles is the $\cnone\sim\wtil B$ 
for $\nmess=1$ or the $\slepr$ (more precisely,
the $\stauone$) for $\nmess\geq2$. Our numerical results are obtained
by evolving from $M_m=1.1\Lambda$. 

As for the other parameters of low-energy supersymmetry, the
$A$-parameters (describing tri-linear Higgs-squark-squark
and Higgs-slepton-slepton couplings) are suppressed, 
while the generation of the $\mu$ and $B$
parameters (responsible for mixing of the 
Higgs superfields and scalar fields, respectively)
is quite model-dependent (and lies somewhat outside the standard
ansatz of gauge-mediated supersymmetry breaking).  
In the end, the parameters of the minimal GMSB model \cite{dtw} consist of 
the normal 18 Standard Model parameters and three new parameters:
$\Lambda$, $\tanb=v_u/v_d$ (the ratio of the vacuum expectation
values of the neutral members of the Higgs doublet fields, $H_u$ and $H_d$,
responsible for giving mass to up-type and down-type quark fields,
respectively) and ${\rm sign}(\mu)$. This is the result after 
trading in the $B$ and $|\mu|^2$ parameters for $v^2=v_u^2+v_d^2$ and $\tanb$,
and using the observed value of $\mz$ to fix $v$.
For this study, we have taken $\tanb=2$ and sign($\mu)=-1$.
Further, we will consider only $\nmess=1$ models in this paper.

\begin{figure}[ht]
\leavevmode
\epsfxsize=5.5in
\centerline{\epsffile{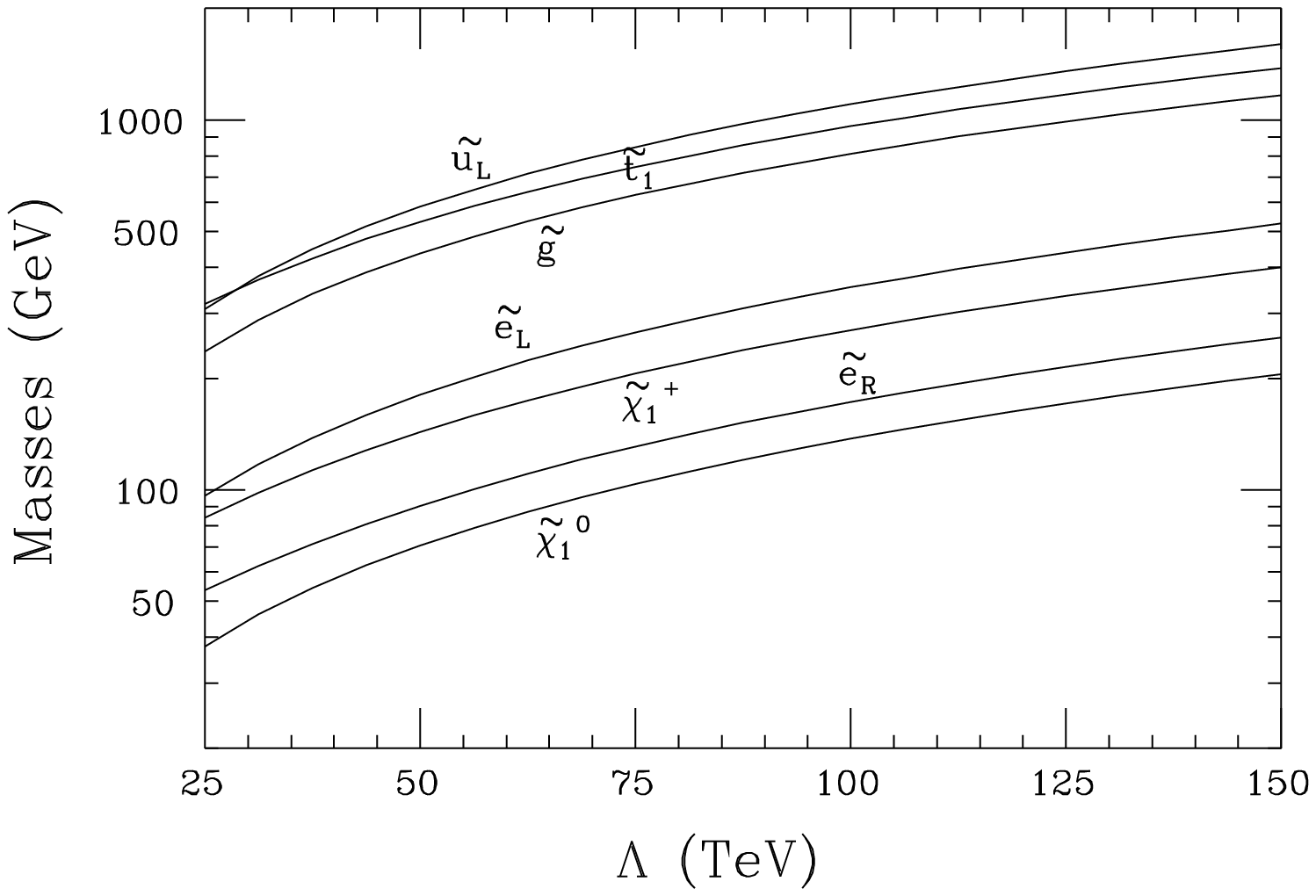}}
\bigskip
\caption{\baselineskip=0pt $\mcnone$, $\mcpone$, $\mselr$, $\msell$, 
$\mgl$, $\mstopone$ and $m_{\wt u_L}$
as functions of $\Lambda$ in the $\nmess=1$ GMSB scenario,
taking $f=g=1$ and $M_m=1.1\Lambda$.}
\label{gmsbmassfig}
\end{figure}

The masses of the sparticles as a function of $\Lambda$, resulting
from the above inputs, are easily summarized. For $\nmess=1$ we have the
following approximate rules of thumb when $\Lambda$ is of order 100 TeV:
\bea
&\mcnone\sim 1.35\gev\times \Lambda(\tev)\,,\quad \mcpone\sim
2.7\times\Lambda(\tev)\,,\quad \mgl\sim 8.1\gev\times\Lambda(\tev)\,,&\nonumber\\
&\mslepr\sim 1.7\gev\times\Lambda(\tev)\,,\quad \mslepl\sim
3.5\gev\times\Lambda(\tev)\,,\quad\msq\sim11\gev\times\Lambda(\tev)\,,&
\label{lammassrel}
\eea
where $M_1$ is the
U(1)-gaugino soft-supersymmetry-breaking mass, $\mgl$ the gluino mass,
$\msq$ a typical $u$ or $d$ squark mass,
and $\mslepr,\mslepl$ are the masses of the right- and left-handed
sleptons (all evaluated at the appropriate $Q\leq 1\tev$).
The masses of various sparticles in the case of $\nmess=1$
are plotted as a function of $\Lambda$ (taking $f=g=1$) in
Fig.~\ref{gmsbmassfig}. 

The range of $\Lambda$ of interest in this model is easily
determined using the dependence of the sparticle masses on $\Lambda$.
An approximate upper limit is set by requiring
that all superpartner masses be $\lsim 1.5\tev$.
An approximate lower limit comes from requiring that the right-handed
slepton be heavier than the rough lower limit of $\sim 70-80\gev$
from current LEP-2 data \cite{aleph} (see later discussion).
From Fig.~\ref{gmsbmassfig} and \Eq{lammassrel},
we see that $\Lambda$ values below $50-60\tev$ would yield
too small a value for $\mslepr$. For $\Lambda$ values above
about $150\tev$ the squarks and gluino would have masses of order
$1.5\tev$, which is somewhat beyond the range for which the
model would provide a comfortable solution to the naturalness problem.

Assuming that supersymmetry is spontaneously broken 
in the hidden sector~\cite{fayet,others}, a massless Goldstone
fermion, the {\it goldstino}, arises.  Its coupling to a particle and
its superpartner is fixed by the supersymmetric Goldberger-Trieman relation:
\beq
\call_{\rm int}=-{1\over F}\,j^{\mu\alpha}\partial_\mu \wt G_\alpha+{\rm
h.c.}\,,
\label{gtrel}
\eeq
where $j^{\mu\alpha}$ is the supercurrent, which depends bilinearly on
all the fermion--boson superpartner pairs of the theory and
$\wt G_\alpha$ is the spin-1/2 goldstino field (with spinor index
$\alpha$).
$\sqrt{F}$ is the scale at which supersymmetry-breaking occurs
in the hidden sector.
When gravitational effects are included, the goldstino is ``absorbed''
by the {\it gravitino} ($\gtino $),
the spin-3/2 partner of the graviton.
By this super-Higgs mechanism, the goldstino is removed from the
physical spectrum and the gravitino acquires a mass ($m_{3/2}$).
In models where the gravitino mass is generated at tree-level one finds
\begin{equation}
m_{3/2}={F\over \sqrt 3\mpl}\,,
\label{mgravitino}
\end{equation}
where $\mplanck$ is the
reduced Planck scale [$\mplanck=(8\pi G_N)^{-1/2}\sim 2.4\times 10^{18}\gev$].
The helicity $\pm\half$ components of
the gravitino behave approximately like the goldstino, whose
couplings to particles and their superpartners are determined by
\Eq{gtrel}.  In contrast, the helicity $\pm\threehalf$ components of
the gravitino couple with gravitational strength to particles and
their superpartners, and thus can be neglected. It is convenient
to rewrite \Eq{mgravitino}
for the gravitino mass as follows:
\begin{equation}
m_{3/2}={F\over \sqrt 3\mpl}\sim 2.5
\left({\sqrt{F}\over 100\tev}\right)^2~{\rm eV}\,.
\label{mgform}
\end{equation}

In the models we consider, the gravitino will be the lightest
supersymmetric partner. The $\cnone$ and $\stauone$ will be substantially
heavier; whichever of these two particle is the lighter will be called
the next-to-lightest supersymmetric particle (NLSP), and will decay
to the gravitino and its SM partner.
If R-parity is conserved, as we assume,
the gravitino will then be stable and thus will be a candidate for dark
matter.  If the gravitino LSP is too heavy ($\mgtino\gsim
{\rm few~keV}$) its relic density will overclose the universe in
most cosmological scenarios~\cite{coslimit,murayamabigf}.  
From \Eq{mgform}, this means that $\sqrt F$
values above roughly $3000\tev$ are disfavored. On the other hand,
there is no particular problem with
very light gravitinos (eV masses). Although these
will not contribute significantly to the total mass density of the
universe, it may turn out that the main component of the dark matter
has another source.  Some examples are:
the QCD axion, its supersymmetric partner (the axino \cite{axino})
or the lightest stable particle in the GMSB messenger sector
\cite{taohan}.

The scale of supersymmetry breaking, 
$\sqrt F$, is a very crucial parameter for the phenomenology
of GMSB models. However,
it is highly model-dependent.  In hidden-sector models, values
of $\sqrt{F}\sim 10^3$--$10^4$~TeV are required in order to generate
sufficiently large supersymmetry-breaking in the sector of MSSM
fields \cite{gunionbigf,murayamabigf}.
In particular, one can derive \cite{gunionbigf} the approximate inequality
\beq
\left({\sqrt F\over 2000\tev}\right)\geq f \left({\mslepr\over 100\gev}\right)
\left({1\over 27.5\sqrt{\nmess}}\right)\,,
\label{bigfeq}
\eeq
where $f=F/F_S\sim \left({g_m^2\over 16\pi^2}\right)^{-2}\sim
2.5\times 10^4/g_m^4$, with
$g_m$ being the coupling of the gauge group
responsible for communicating (via two-loop diagrams)
supersymmetry breaking from the
hidden sector to the messenger sector. Perturbativity would require
$g_m\lsim 1$.
For $\mslepr\geq 75-80\gev$ (the rough LEP-2 limit)
and $f=2.5\times10^4$, \Eq{bigfeq} yields
$\sqrt F\geq 10^4\tev$ ($\sqrt F\geq 5000\tev$)
for $\nmess=1$ ($\nmess=4$). Thus, even allowing for the roughly factor
of 2 or 3 uncertainty in the approximate lower bound, $\sqrt F$ must lie
near its upper limit from cosmology.
In direct-transmission models, the two-loop communication
between the hidden and messenger sectors is eliminated, $f$ 
in \Eq{bigfeq}\ is effectively of order unity, 
and $\sqrt{F}$ can be as low as 100~TeV in phenomenologically viable models.

In what follows, we will focus on the ($\nmess=1$) model in which
the $\cnone$ is the next-to-lightest supersymmetric particle.~\footnote{Focus
on this model was originally motivated by the $ee\gam\gam$ event
at the Tevatron~\cite{park}. However, recent Tevatron analyzes
\cite{limitsoneegg} rule out most, if not all, of the parameter
space for which this model could have explained this event.}
We will employ the specific input boundary conditions given earlier.
With these inputs, the model is completely specified in terms of 
just two independent parameters,
$\Lambda$ and $\sqrt F$. As explained above,
$\Lambda$ determines the  gaugino and
sfermion masses, while $\sqrt F$ determines the properties of
the gravitino and (see below) the lifetime of the $\cnone$.
For $\sqrt F$ in the $100$--$3000\tev$ range, the small size of the couplings,
\Eq{gtrel}, imply that all the sparticles other than the $\cnone$
undergo chain decay down to the $\cnone$. The $\cnone$
finally decays to a two-body final state containing the $\gtino$: 
$\cnone\to\gam\gtino$ is the only allowed two-body mode
for $\Lambda\lsim 80\tev$, and this mode
remains dominant out to the highest $\Lambda$ values
considered.~\footnote{The other two-body modes
that become kinematically allowed at high $\Lambda$
are $\cnone\to Z\gtino$ and $\cnone\to \hl\gtino$. However,
for a $\cnone$ that is nearly pure bino, the $\cnone\to Z\gtino$
width is suppressed relative to that for $\cnone\to \gam\gtino$ by a factor
of $[\sin\theta_W/\cos\theta_W]^2$ and the $\cnone\to\hl\gtino$
partial width is proportional to the square of the
very small higgsino component of the $\cnone$.}
The $c\tau$ for $\cnone\to\gam\gtino$ decay takes the form:
\begin{equation}
(c\tau)_{\cnone=\wtil B\to \gam \gtino}\sim 130
\left({100\gev\over \mcnone}\right)^5
\left({\sqrt F\over 100\tev}\right)^4\mu {\rm m}
\label{ctauform}
\end{equation}
If $\sqrt F\sim 3000\tev$ (the upper limit from cosmology),
then $c\tau\sim 100$m for $\mcnone=100\gev$; 
$\sqrt F\sim 100\tev$ implies a short but vertexable decay length. 
Thus, the signatures for this GMSB model are crucially dependent
on $\sqrt F$. In particular, for large $f= F/F_S$,
and hence large $\sqrt F$, the neutralino decay is most naturally characterized
by a substantial decay length, possibly of order tens to hundreds of meters.
Events in which one or more of the neutralinos travels
part way through the detector,
and then decays, can be a substantial fraction of the total,
as can events in which all the neutralinos exit the detector
before decaying. Thus, at the very least, 
it is highly relevant to assess how our ability to discover
supersymmetry changes as a function of $\sqrt F$. In the following sections
of this paper we will focus on this issue 
in the context of the D0 detector at the Tevatron collider.

Before proceeding with the Tevatron analysis, let us return
to the question of what
the current LEP-2 limits are upon the simple GMSB
model we are considering.  The latest limits \cite{aleph} including
$\rts=183\gev$ running are quite significant, at least in the 
limits of either small $\sqrt F$ (for which $\cnone\to \gam\gtino$
is prompt) or large $\sqrt F$ (for which the $\cnone$ is invisible).
In the small $\sqrt F$/prompt-photon limit, slepton mass
limits of $\gsim 75-80\gev$ arise from looking
for $\slepr\slepr$ events where $\slepr\to\ell\cnone\to\ell\gam\gtino$.
Further, $\mcnone\gsim 80\gev$
is required by the absence of excess $\gam\gam+\etmiss$ events.
In combination, these require $\Lambda\gsim 60\tev$ (see
Fig.~\ref{gmsbmassfig}). In the large $\sqrt F$/invisible-$\cnone$
limit, the normal SUSY search for $\slepr\slepr$ with 
$\slepr\to\ell\cnone$ 
implies $\mslepr\gsim 80\gev$, implying $\Lambda\gsim 55\tev$.
The case of intermediate $\sqrt F$ has not yet been specifically analyzed,
but we would anticipate that (conservatively)
$\Lambda\leq 50-55\tev$ would be excluded by such an analysis.
Thus, as stated earlier, we shall focus our analysis on the $\Lambda\geq
50-60\tev$ region.

Finally, we wish to caution that 
many alternative GMSB models have now been constructed.
In particular, the various direct-transmission models 
are very different from one another and need not have
the simple $M_m$-scale boundary conditions of Eqs.~(\ref{gmsbmi})
and (\ref{gmsbmsq}). 
The relations between sparticle and gravitino masses vary tremendously.
As an extreme example, in the model of Ref.~\cite{raby} the
gluino is the LSP and the gravitino is sufficiently heavy as to be
phenomenologically irrelevant.

The remainder of the paper is organized as follows.
In Sec.~2, we describe in detail the 
phenomenology of the $\cnone$-LSP
model and give results for all the signals considered.
In particular, we describe in detail the delayed $\cnone$ decay
signals and illustrate the regions of model
parameter space for which they would allow detection of a supersymmetry
signal.  Sec.~3 presents discussion and conclusions.

\section{Collider phenomenology}

In the Introduction, we emphasized that our ability to detect supersymmetry
at the Tevatron will depend substantially on $\sqrt F$.
The case where $\sqrt F$ is sufficiently small
that all $\cnone$ decays are prompt received early
attention~\cite{ddrt,akkmm,dtw,bbct,bmpz}. Of these references,
only \cite{bbct} performs a quantitative 
Tevatron collider study with the full boundary 
condition constraints of Eqs.~(\ref{gmsbmi}) and (\ref{gmsbmsq}).
The results of Ref.~\cite{bbct} were confirmed in Ref.~\cite{chenguniongampub}.
In these latter analyzes, the prompt-decay
hadron collider signal for supersymmetry
considered was events containing two or more isolated photons, deriving
from decay of two or more neutralinos, plus missing energy
(and possibly other particles).
The case where $\sqrt F$ is very large
and most $\cnone$ decays occur outside the detector was first analyzed
in Ref.~\cite{chenguniongampub}; it is equivalent
to conventional supersymmetry phenomenology (in which, for example,
one looks for jets plus missing energy or three leptons plus missing energy) 
with the constraints
implied by Eqs.~(\ref{gmsbmi}) and (\ref{gmsbmsq}) among
the sparticle masses.  For moderate $\sqrt F$, it is most 
appropriate to look for signals that would be characteristic of
events in which the $\cnone$ decay is delayed, but still occurs
within the detector (or within a region that could be probed
by a simple addition to the detector). In Ref.~\cite{chenguniongampub}
one signal of this type was examined in the
context of the D0 detector: namely, 
the `outer-hadronic-calorimeter' (OHC) signal
that results when the delayed decay takes
place in one of the outermost D0 hadronic calorimeter cells.
Here, we will consider in addition, two other delayed-decay 
signals. The first is the `impact parameter' ($b$) signal. 
If the (transverse) impact parameter of a photon appearing
in the electromagnetic calorimeter could be shown to be non-zero,
this would constitute a signal for a delayed decay.  The extent to which
the resolution for such an impact parameter measurement by the
CDF and D0 detectors is adequate for this signal to be useful is
an important issue. The second is the `roof-array' (RA) signal.
For large $\sqrt F$, many $\cnone$ decays will occur outside
the muon chambers.  A two-layer detector placed on the roof
of the D0 detector building could pick up a portion of
such delayed photons and possibly provide a dramatic signal.
In fact, we will demonstrate that
sensitivity to GMSB models with substantial $\cnone$ decay length will
be maximized by implementing all three (OHC, $b$, RA) delayed-decay signals.

We will now outline the results we have obtained for these signals
and the associated procedures.
Events were generated simultaneously
for all \susy\ production mechanisms by modifying 
ISASUGRA/ISAJET~\cite{isajet} to incorporate the GMSB boundary conditions,
and then forcing delayed $\cnone$ decays according to the predicted $c\tau$.
We used the toy calorimeter simulation package ISAPLT. We simulated
calorimetry covering $|\eta|\leq 4$ with a cell size given by
$\Delta R\equiv \Delta\eta\times\Delta\phi=0.1\times 0.0875$ and took the
hadronic (electromagnetic) calorimeter resolution to be
$0.7/\sqrt E$ ($0.15/\sqrt E$). The D0 electromagnetic calorimeter
was simplified to a thin cylinder with radius $r=1$m
and length $-2{\rm m}\leq z\leq +2$m. Also important to the analysis was
the central outer hadronic D0 calorimeter (OHC), which we approximated
as occupying a hollow solid cylinder defined by $-2{\rm m}\leq z\leq +2$m and 
radial region $2{\rm m}\leq r\leq 2.5{\rm m}$. It is important
that the D0 OHC is segmented in $\Delta R$
and that there are also several layers of inner hadronic calorimeter.

The signals are defined in terms of jets, isolated leptons,
isolated prompt photons and isolated delayed-decay photons 
(\ie\ emerging from $\cnone$ decays occurring with
substantial delay). 
\bit
\item
A jet is defined by requiring
$|\eta_{\rm jet}|<3.5$ and  $\etjet >25\gev$ (for  
$\Delta R_{\rm coal.}=0.5$). 
\item
An isolated lepton is defined by requiring $|\eta_\ell|<2.5$ and
$E_T^{\ell}>20$, $15$, $10\gev$ for the most energetic, 2nd most
energetic and 3rd most energetic lepton, respectively.~\footnote{We never
consider more than three isolated leptons, and all leptons are prompt.}
The criterion for isolation is $E_T(\Delta R\leq 0.3)<4\gev$ (summing over 
all other particles in the cone surrounding the lepton).
\item
A photon is a prompt-photon candidate
if it emerges from a $\cnone$ decay that
occurs before the $\cnone$ has reached the electromagnetic calorimeter.
The $(\eta_\gamma,\phi_\gamma)$ of such a photon
is defined by the direction
of the vector pointing from the interaction point to the point at which it
hits the electromagnetic calorimeter.
(This is generally not the same as the direction of the photon's momentum.)
An isolated prompt photon is defined by requiring
$|\eta_\gamma|<1$ and $E_T^\gamma>12\gev$,
with isolation specified by $E_T(\Delta R\leq 0.4)<4\gev$, summing over 
all other particles in the cone surrounding the point
at which the photon hits the EM calorimeter.
Photons that emerge within the hollow cylinder defined by 
the electromagnetic calorimeter,
but that are not isolated, are merged with hadronic jets as appropriate.
\item
An isolated delayed-decay OHC signal 
photon is defined as one that emerges from a
$\cnone\to \gamma \gtino$ decay that takes place inside an OHC cell
and that has $E_T^\gamma>15\gev$, with isolation specified
by $E_T(\Delta R\leq 0.5)<5\gev$, summing over all other particles
in the cone surrounding the location of the $\cnone$ decay.
This implies, in particular, that
there should be very little energy deposited in
the inner hadronic calorimeter cells in the same $(\eta,\phi)$ location
as the OHC cell in which the $\cnone$ decay occurs.
\item
An isolated delayed-decay impact-parameter
signal photon is defined as one that emerges from a
$\cnone\to \gamma \gtino$ decay (occurring before the $\cnone$
has exited the tracking region) and that passes through the
central electromagnetic calorimeter (with the dimensions 
and physical location specified earlier). 
The photon is required to have
$E_T^\gamma>15\gev$. No isolation requirement is imposed.
Nor is there any specific requirement on the apparent rapidity (\ie\
as measured by the location at which it passes through the electromagnetic
calorimeter).  
\item
A delayed-decay roof-array 
signal photon is defined as one that emerges from a
$\cnone\to \gamma \gtino$ decay that occurs
at a vertical distance in the range $6.5{\rm m}<z<16.5$m from the interaction
point and has $E_T^\gamma>15\gev$. 
This means that the $\cnone$ will have decayed
past the muon detector system but below the roof of the D0 detector building.
The delayed-decay photon is required to pass through a 28m$\times$38m
rectangle centered at 16.5m above the interaction point. This is 
the approximate size of roof array that could be positioned so that 
no part of the array would have a large amount of material between it
and the 6.5m height referred to earlier. No further isolation
requirements or other cuts are placed on the photon.
\eit

We now give the specific requirements for each type of signal. First,
for all signals, events are retained only if at least one of several sets 
of reasonable trigger requirements are satisfied. 
The triggers considered were the following.
\begin{enumerate}
\item
$\etmiss>35\gev$.
\item
four or more jets with $E_T^{\rm jet}>15\gev$.
\item
(a) two or more jets with $E_T^{\rm jet}>30\gev$;
(b) $\etmiss>40\gev$.
\item
(a) one or more leptons with $E_T^\ell>15\gev$, $|\eta_{\ell}|<2.5$ 
and isolation defined by $E_T(\Delta R<0.3)<5\gev$; (b) $\etmiss>15\gev$.
\item
(a) two or more leptons with $E_T^\ell>15\gev$, $|\eta_{\ell}|<2.5$ 
and isolation defined by $E_T(\Delta R<0.3)<5\gev$;
(b) $\etmiss>10\gev$.
\item
two or more photons with $E_\gam>12\gev$, $|\eta_\gam|>1$ and isolation
specified by $E_T(\Delta R<0.4)<4\gev$.
\end{enumerate}
The additional signal-specific cuts are itemized below.
\bit
\item The standard jets-plus-missing-energy signal.
\eit

We employ D0 cuts
\cite{dzero}: 
(a) $n({\rm jets})\geq 3$ --- labelled $k=1,2,3$ according to decreasing $E_T$;
(b) no isolated ($E_T^{\rm had.}(\Delta R\leq 0.3)<5\gev$)
leptons with $E_T>15\gev$;
(c) $\etmiss>75\gev$; 
(d) $0.1<\Delta\phi(\etmiss,j_k)<\pi-0.1$ and
$\sqrt{(\Delta\phi(\etmiss,j_1)-\pi)^2+(\Delta\phi(\etmiss,j_2))^2}>0.5$.
The background cross section level has been estimated by D0 to be
$\sigma_B=2.35\pb$. The signal will be deemed observable if: (i)
there are at least 5 signal events; (ii) $\sigma_S/\sigma_B>0.2$; and (iii)
$N_S/\sqrt{N_B}>5$, where $N_S$ and $N_B$ are the numbers of signal
and background events.

\bit
\item The tri-lepton signal.
\eit

We employ the cuts of Ref.~\cite{baertri}: 
three isolated leptons, $n({\rm jets})=0$, $\etmiss>25\gev$
and $|M(\ell_i\ell_j)-\mz|>8\gev$ ($i\neq j$).
For these cuts, the cross section for the
sum of all backgrounds to the tri-lepton signal is $\sigma_B=0.2\fb$.

\bit
\item The tri-lepton-plus-prompt-photon signal.
\eit

We impose the same cuts as for the tri-lepton
signal, but require in addition
that there be at least one prompt photon (as defined above) present.
We shall presume that the background to this signal will be negligible.
This seems reasonable given the already quite small background to the
simple tri-lepton signal.

\bit
\item
The two-prompt-photon-plus-missing-energy signal.
\eit

Following Ref.~\cite{bbct}, 
we require: (a) at least two isolated prompt photons (as defined above); 
and (b) $\etmiss>\etmin$.
Detection efficiency of 80\% (100\%) is assumed if 
$E_T^\gamma<25\gev$ ($>25\gev$). This signal should be completely
free of background provided $\etmin$ is adjusted appropriately
as a function of luminosity.
The signal will be deemed observable if there are 5 or more events.

\bit
\item 
The delayed-neutralino-decay signal resulting when the $\cnone$ decay takes
place in the outermost hadronic calorimeter cell of the D0 detector.
\eit

In an event in which
a $\cnone\to \gamma \gtino$ decay occurs inside one
of the outer hadronic calorimeter (OHC) cells, the $\gamma$ will
deposit all its energy in the cell. By demanding substantial $\gamma$ energy 
and isolation (precise requirements were given earlier)
for this deposit, along with other criteria, backgrounds
can be made small.
\footnote{Our additional criteria will be chosen so that the cross section for
producing an isolated energetic long-lived kaon in association
therewith is small. Further, any such kaon will interact
strongly and is almost certain to be absorbed before reaching the OHC.}
More specifically, we require that the event fall into
one of three classes defined by the following sets of requirements:
\begin{enumerate} 
\item
(a) $n({\rm jets})\geq 3$; 
(b) at least one delayed-decay OHC photon;
(c) $\etmiss>\etmin$. 
\item 
(a) any number of jets;
(b) two or more delayed-decay OHC photons;
(c) $\etmiss>\etmin$. 
\item 
(a) any number of jets;
(b) at least one prompt photon;
(c) at least one delayed-decay OHC photon;
(d) $\etmiss>\etmin$. 
\end{enumerate}
In the absence of the needed
detector-specific study, we will assume zero background to 1.+2.+3.
for the same $\etmin$ values that will be employed
for the prompt-$2\gamma$ signal. 
Observability of the delayed-decay OHC signal is assumed if the number of
events for 1.+2.+3. is 5 or more. Of the above three sub-signals, 1.
provides by far the largest event rates in the moderate $\sqrt F$ region.

\bit
\item  The impact-parameter signal.
\eit

We require $n({\rm jets})>3$, $\etmiss>\etmin$ and one or more delayed-decay
impact-parameter photons (as defined above) 
with transverse impact parameter $b$
larger than 2cm.  The D0 detector in Run-II will have a pre-shower installed
and is expected \cite{landsberg} to achieve a $1\sigma$ resolution of about
$0.2$cm in the transverse impact parameter. Thus, our 2cm requirement
amounts to a $10\sigma$ deviation from zero impact parameter.

\bit
\item The roof-array signal.
\eit

We require $n({\rm jets})>3$, $\etmiss>\etmin$ and one or more delayed-decay
roof-array photons. If there is directionality or timing for
the roof array, additional cuts that ensure tiny backgrounds
could be incorporated without significant losses in the signal event rate.
However, such additional cuts are not imposed in this study.

We will now summarize our results for all these channels.
A limited subset of these results
appeared in Ref.~\cite{chenguniongampub}. In the present paper,
our discussion will focus on
the luminosities expected in Run-II and at TeV33.~\footnote{We will
not discuss the delayed-decay signals for Run-I luminosity here.
The OHC signals
give only a marginal increase in the range of parameter space for
which a supersymmetry signal might emerge for the low $L\sim 100\pbi$
integrated luminosity accumulated in that run. 
Only if the stringent cuts we 
employ for the delayed-decay OHC signals could be weakened without
encountering backgrounds would present data allow elimination
of a significant additional portion of parameter space. Background
studies are now in progress~\cite{mani}. The impact-parameter
and roof-array signals cannot be implemented for the Run-I data set.}
We begin with the jets-plus-missing-energy and tri-lepton signals.
\bit
\item The jets-plus-missing-energy-signal.

Employing the cuts specified above,
for $L=0.1\fbi$ (Run-I), we found in Ref.~\cite{chenguniongampub}
that this signal was viable for any $\sqrt F$ if $\Lambda\lsim 25\tev$.
For both $L=2\fbi$ and $L=30\fbi$, 
the signal is viable for any $\sqrt F$ if $\Lambda\lsim 30\tev$.
Given the lower bound of $\sim 55\tev$ coming from the lower bound
on $\mslepr$, it would appear that the jets-plus-missing-energy
signal will not be useful at the Tevatron.

\item The tri-lepton signal.

\begin{figure}[h]
\leavevmode
\epsfxsize=5.5in
\centerline{\epsffile{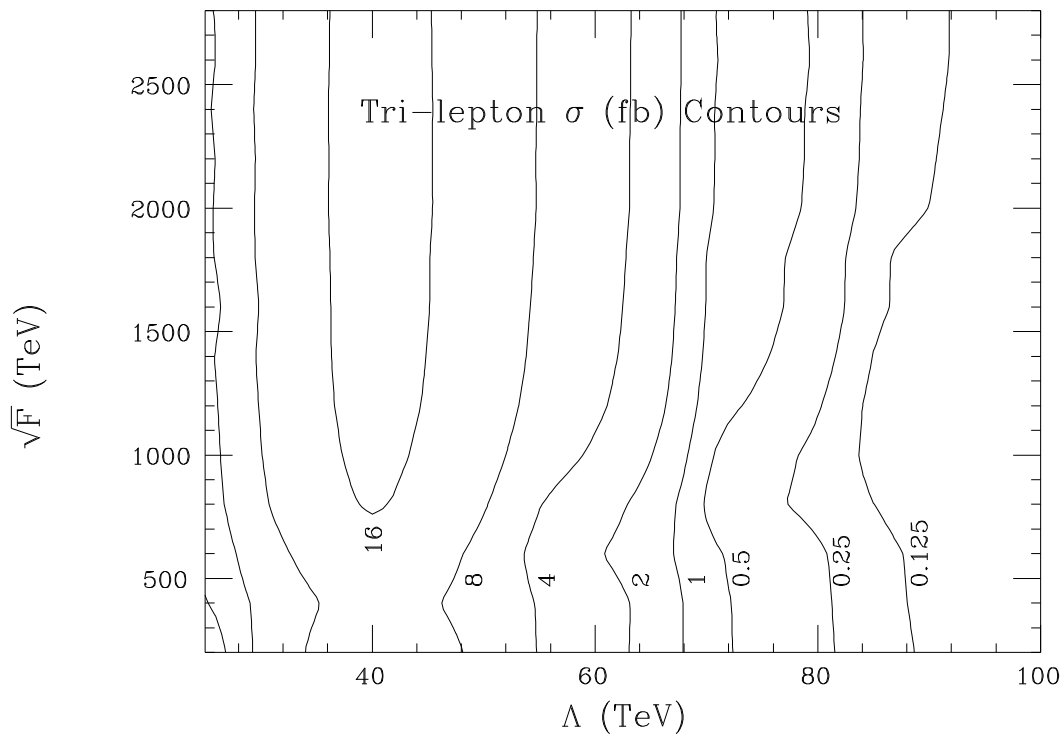}}
\bigskip
\caption{\baselineskip=0pt 
We present cross section contours in fb units
for the tri-lepton signal (as defined in the text).}
\label{trileptonfig}
\end{figure}

Contours of constant cross section for this signal are presented
in Fig.~\ref{trileptonfig}.  Comparing to the background cross section,
$\sigma_B=0.2\fb$, we see that in order to achieve $S/\sqrt B=5$,
we require a signal cross section of $\sigma=8.4\fb$
at $L=0.1\fbi$ (Run-I) $\sigma=1.6\fb$ at $L=2\fbi$ (Run-II) 
and a signal cross section of $\sigma=0.41\fb$ at $L=30\fbi$ (Tev33), 
respectively. Fig.~\ref{trileptonfig}
then shows that the parameter space for which
supersymmetry can be discovered using this signal
extends to $\Lambda\sim 50\tev$ for $L=0.1\fbi$,
to $\Lambda\sim 65\tev$ for $L=2\fbi$ and to $\Lambda\sim 75\tev$
for $L=30\fbi$.  Thus, the tri-lepton discovery mode will
allow supersymmetry detection out to substantially higher $\Lambda$
than does the jets plus missing energy mode.  (This is to be 
expected since the tri-lepton signal tends to be stronger than
the jets-plus-missing-energy signal at Tevatron energies~\cite{baertri}.)
In particular, it can probe above the $\sim 55\tev$ lower
bound on $\Lambda$ coming from LEP-2 limits on $\mslepr$.

\eit
For both these signals, we should note that if $\sqrt F$ is small enough
then, in association with the jets or leptons,
one will detect prompt photons (coming from
the decays of the $\cnone$'s that are inevitably present).
The prompt-two-photon signal to be discussed below will also be
in evidence.  There will then be little doubt that nature
has chosen a GMSB model rather than a SUGRA~\footnote{As usual,
the SUGRA acronym refers to models in which supersymmetry
breaking in a hidden sector is transmitted to the supersymmetric 
partners of the SM particles by gravitational interactions rather
than gauge interactions. The gravitino in such models is heavy and
an LSP $\cnone$ is stable in the absence of R-parity violation.}
model with similar boundary conditions for the 
gaugino and sfermion soft-supersymmetry-breaking masses.
However, if $\sqrt F$
is large there will be no prompt photons and no hint for the GMSB
character of the model from these two signals. To illustrate these
points, we turn to the prompt-photon signals.

\bit
\item The tri-lepton-plus-prompt-photon signal.

\begin{figure}[h]
\leavevmode
\epsfxsize=5.5in
\centerline{\epsffile{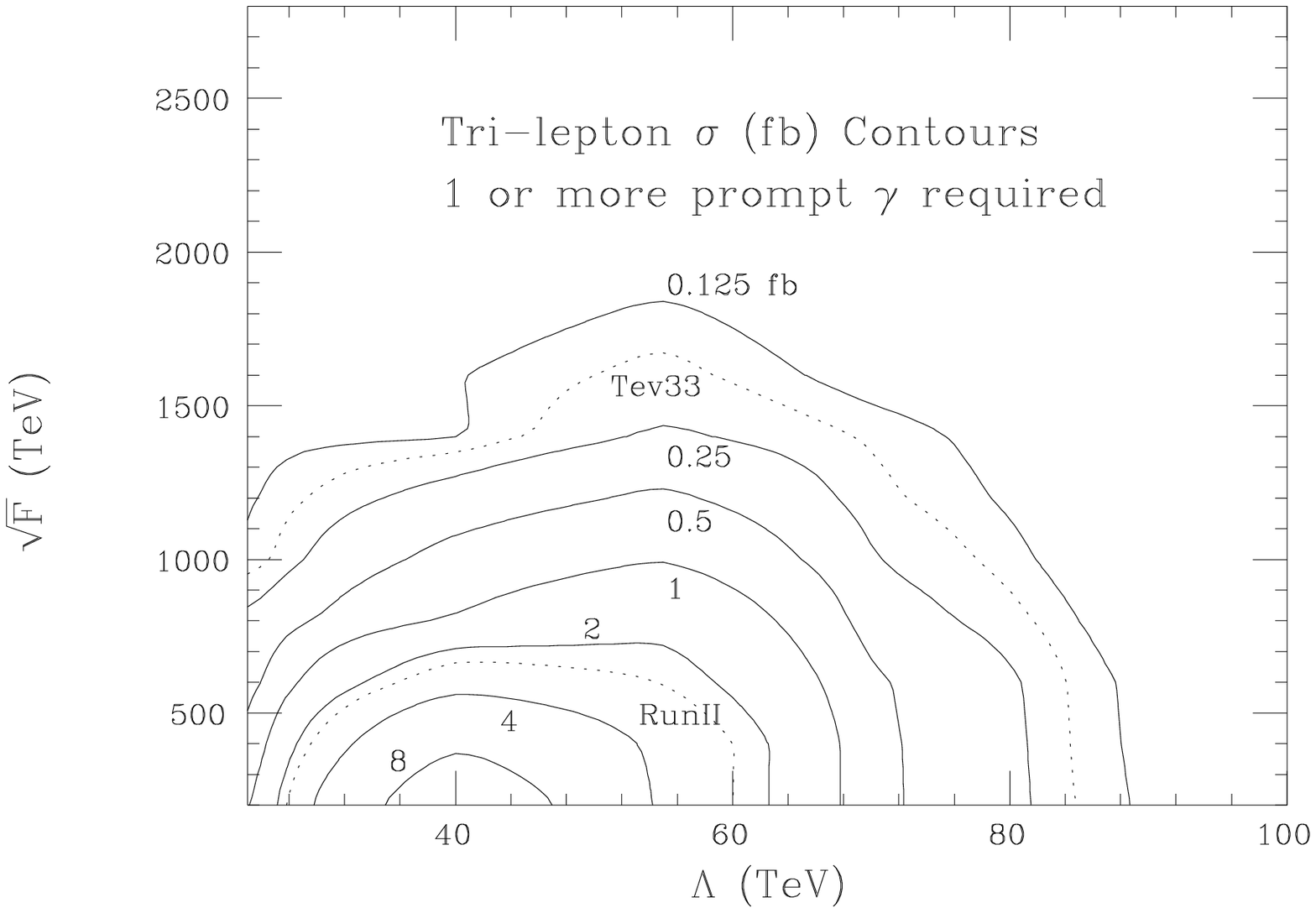}}
\bigskip
\caption{\baselineskip=0pt 
We present cross section contours in fb units
for the tri-lepton-plus-prompt-photon signal (as defined in the text).
In addition to the cuts for the tri-lepton signal, we impose the
requirement that there be at least one prompt photon present. 
The solid contours are the same ones presented in Fig.~\ref{trileptonfig}.
The dotted contours correspond to 5 events at Tev33 ($L=30\fbi$)
and 5 events in Run-II ($L=2\fbi$).}
\label{trileptonpromptfig}
\end{figure}

Contours of constant cross section for this signal are presented
in Fig.~\ref{trileptonpromptfig}.  The solid contours are the same as
plotted in Fig.~\ref{trileptonfig}.  
The dotted contours correspond to 5 events at Tev33 ($\sigma=0.16\fb$
at $L=30\fbi$) and 5 events in Run-II ($\sigma=2.5\fb$ at $L=2\fbi$).
Five events should be adequate for discovery given the negligible background
expected. We observe that prompt photons will be seen in association
with the tri-lepton signal only for $\sqrt F\lsim 600\tev$ at $L=2\fbi$ and
only for $\sqrt F\lsim 1600\tev$ at $L=30\fbi$.

\eit

\begin{figure}[p]
\leavevmode
\epsfxsize=5.5in
\centerline{\epsffile{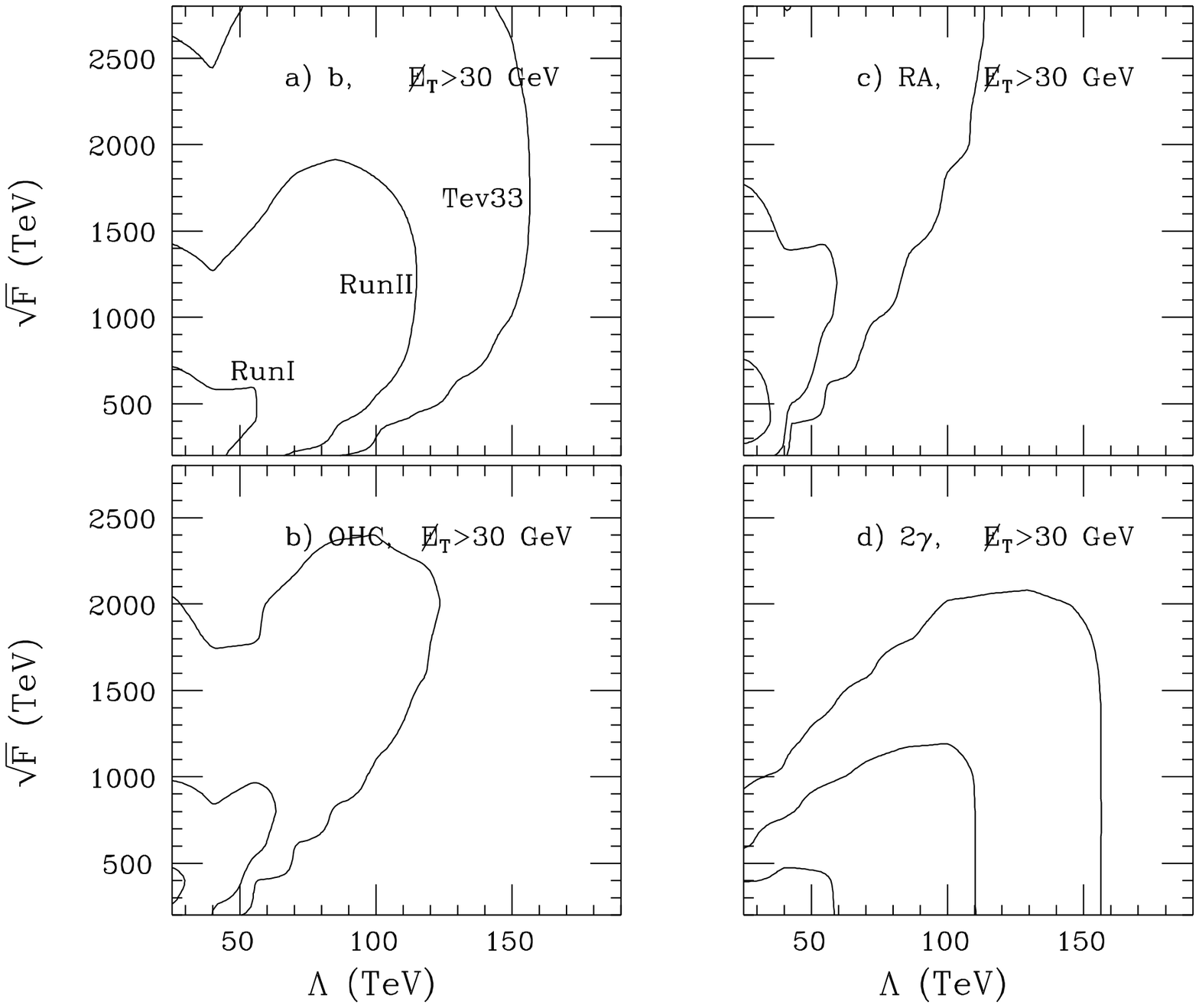}}
\bigskip
\caption{\baselineskip=0pt 
We present cross section contours in the $(\protect\sqrt F,\Lambda)$
parameter space for the (a) impact-parameter ($b$), (b) outer-hadronic (OHC),
(c) roof-array (RA), and (d) prompt-two-photon ($2\gam$) signals.
Contours are given at $\sigma=0.16$, 2.5, and $50\fb$
(the outermost contour corresponding to the smallest cross section, \etc).
We note that $\sigma=0.16,2.5,50\fb$ corresponds to 
5 events at $L=30,2,0.1\fbi$,
respectively. Hence the labels Tev33, Run-II and Run-I, respectively.
In this figure we have taken $\etmin=30\gev$ for all
signals. Other jet and photon requirements are as specified in
the text.}
\label{contoursfullfigiii}
\end{figure}

\begin{figure}[p]
\leavevmode
\epsfxsize=5.5in
\centerline{\epsffile{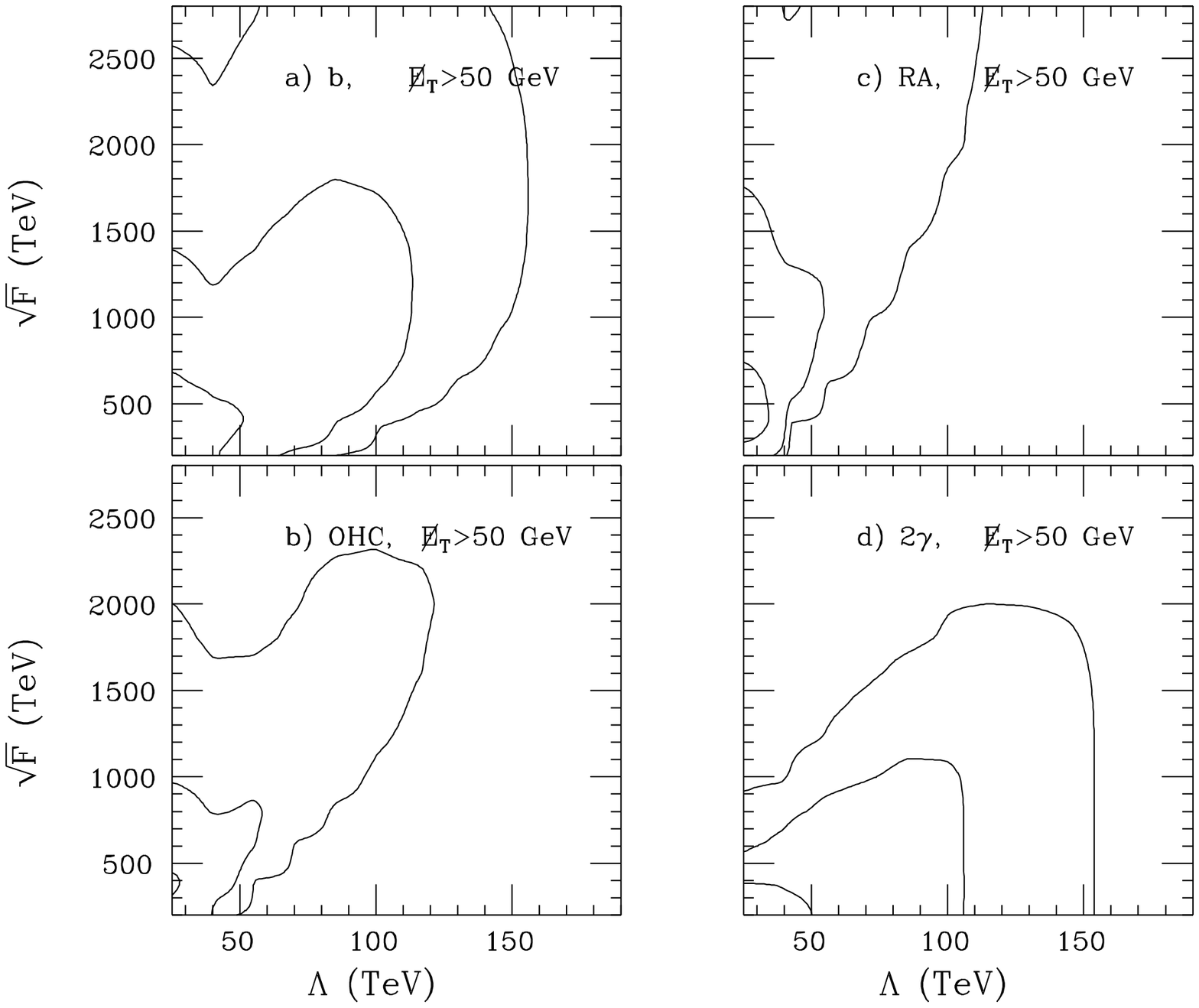}}
\bigskip
\caption{\baselineskip=0pt 
As in Fig.~\ref{contoursfullfigiii}, but for $\etmin=50\gev$.}
\label{contoursfullfigv}
\end{figure}

\begin{figure}[p]
\leavevmode
\epsfxsize=5.5in
\centerline{\epsffile{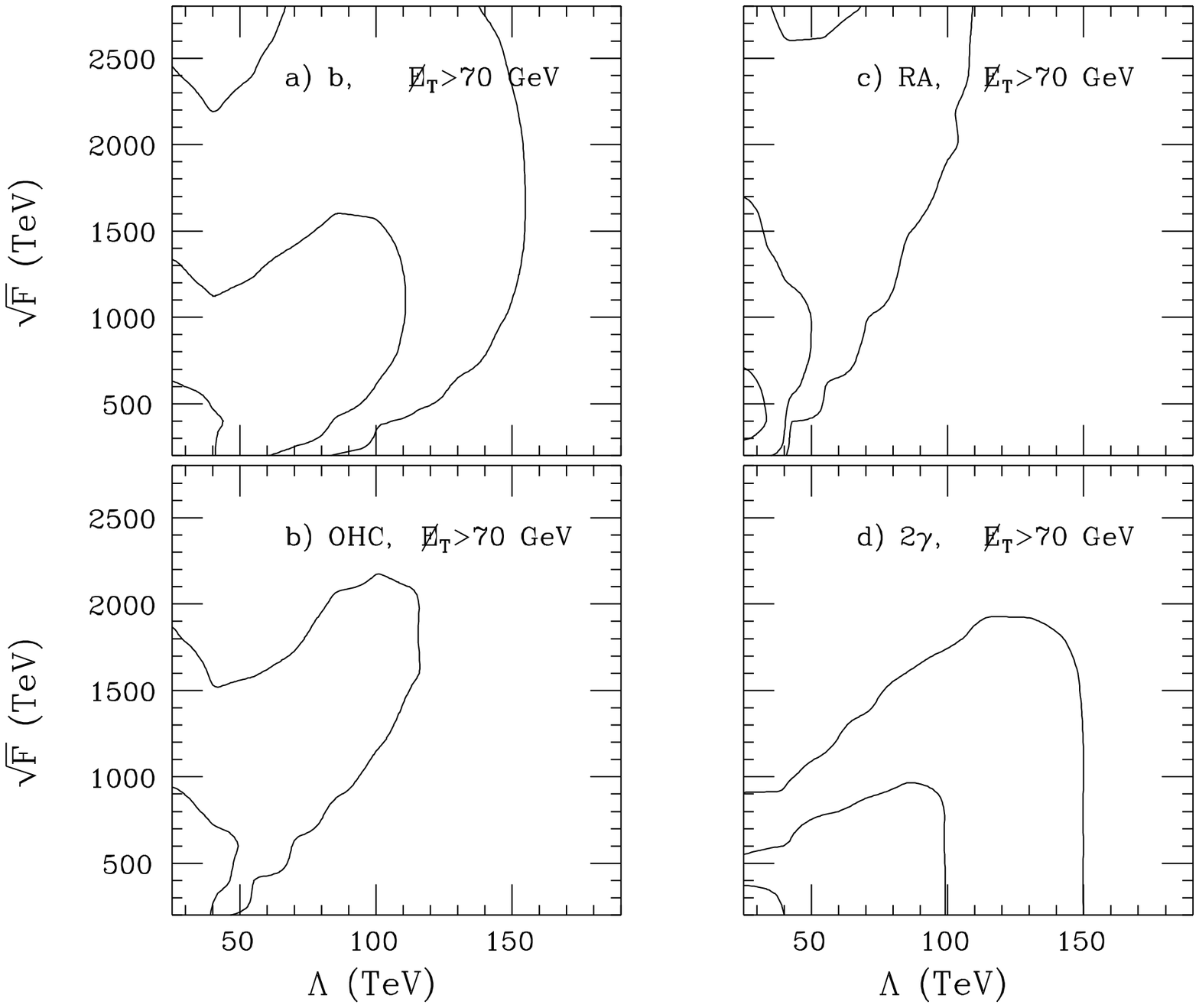}}
\bigskip
\caption{\baselineskip=0pt 
As in Fig.~\ref{contoursfullfigiii}, but for $\etmin=70\gev$.}
\label{contoursfullfigvii}
\end{figure}

We now turn to the impact-parameter ($b$), outer-hadronic-calorimeter (OHC),
roof-array (RA) and prompt-two-photon ($2\gam$) signals.
In all these cases, we shall presume that our cuts can be chosen
so as to eliminate any background. In particular, we will
consider cross sections and 
event rates for the cuts stated earlier with various choices of $\etmin$.
A complete summary of our simulations is contained in
Figs.~\ref{contoursfullfigiii}, \ref{contoursfullfigv},
and~\ref{contoursfullfigvii},
where we give cross section contours for the above signals in the
$(\sqrt F,\Lambda)$ parameter space for $\etmin=30$, 50 and $70\gev$,
respectively.

We now discuss the implications of the results contained in
Figs.~\ref{contoursfullfigiii}, \ref{contoursfullfigv},
and~\ref{contoursfullfigvii}.

\bit
\item The prompt-$2\gam$ signal.

Based on the background results of \cite{d0gamgam} (see also
\cite{cdfgamgam}) and eye-ball extrapolations thereof, we estimate
that our jet requirements and photon energy/isolation requirements
are such that $\etmin=30$, 50, $70\gev$ 
is sufficient to eliminate all background
for luminosities of $L=100\pbi$, $2\fbi$, $30\fbi$, respectively. 
Thus, to ascertain the discovery reach
of this channel for $L=100\pbi$, one refers to the $\sigma=50\fb$ contour
of Fig.~\ref{contoursfullfigiii}; for $L=2\fbi$, one refers to the
$\sigma=2.5\fb$ contour of Fig.~\ref{contoursfullfigv}; and,
for $L=30\fbi$, one refers to the $\sigma=0.16\fb$ contour of
Fig.~\ref{contoursfullfigvii}. From these contours, we conclude that
the signal is viable in a region described as follows.
For $L=2\fbi$, the region extends from small $\sqrt F$ up to $\sqrt F\sim
600\tev$ if $\Lambda\lsim 35\tev$, increasing to $\sqrt F\sim 1000\tev$
at $\Lambda\sim 80-100\tev$; 
the signal is not viable beyond $\Lambda\sim 100\tev$.
For $L=30\fbi$, the region extends from small $\sqrt F$ up to $\sqrt F\sim
900\tev$ if $\Lambda\lsim 35\tev$, increasing to $\sqrt F\lsim 1000\tev$
at $\Lambda\sim 120-130\tev$; 
the signal is not viable beyond $\Lambda\sim 150\tev$.

\item The OHC signal(s).

We provisionally adopt the same $\etmin$ cuts of $\etmin=30$, 50, $70\gev$
for $L=100\pbi$, $2\fbi$, $30\fbi$, respectively, as employed for
the prompt-$2\gam$ signal. 
Assuming that 5 events are sufficient (\ie\ that
the background is negligible for our other
strong cuts on jets and photon energy/isolation)
to give an observable signal, we look at the same $\sigma$
contours as a function of $L$ as in the prompt-$2\gam$ case,
with the following results.
For $L=2\fbi$, the OHC signal provides a useful, 
but not enormous, extension 
of the region of parameter space for which supersymmetry can
be detected as 
compared to the portion of parameter space covered by the
prompt-two-photon signal or jets-plus-missing-energy signal.
This extension is mainly in
the $\sqrt F\lsim 800-1000\tev$, $\Lambda\lsim 50\tev$ region.
Of course, we noted earlier that LEP-2 data probably already
excludes this $\Lambda$ region.
For  $L=30\fbi$, the reach of the OHC signal is greatly extended;
supersymmetry detection may be possible via the OHC signal for essentially all
of the $\sqrt F\lsim 1500-2000\tev$, $\Lambda\lsim 120\tev$ portion
of parameter space for which the prompt-two-photon signal 
is not viable. Conversely, supersymmetry
detection at the low-$\sqrt F$ points at large $\Lambda$
for which the OHC signal falls below 5 events would be possible via
the prompt-$2\gam$ channel. In addition, there is a substantial region
in which both these signals are viable. Since the
tri-lepton signal
only extends out to about $\Lambda\lsim 75\tev$ for $L=30\fbi$,
it is apparent that, at the Tev33 luminosity, the prompt-$2\gam$ 
and OHC signals combine to significantly extend to higher 
values of $\Lambda$ the region where supersymmetry can be detected.
In particular, $\Lambda$ values substantially beyond the approximate
limit of $\Lambda\gsim 55\tev$ coming from LEP-2 data can be probed.

We make two further remarks regarding the OHC signals.
\bit 
\item 
Of the three delayed-decay OHC signals, 
the first ($\geq 3$ jets + $\geq 1$ OHC $\gamma$)
is observable (\ie\ yields at least five events) for the entire
$(\sqrt F,\Lambda)$ region where the OHC signals are shown
to be viable.
The second ($\geq 2$ OHC $\gamma$'s) yields $\geq 5$ events 
for only a very small portion of this region.
The third ($\geq 1$ prompt $\gamma$ + $\geq 1$ OHC $\gamma$)
yields $\geq 5$ events for the lower (roughly, half) $\sqrt F$ portion of the
region for which the combined OHC signals are viable.
\item
If simulations eventually
show that the first OHC signal is background-free for smaller $\etmin$ than
employed here (as we are hopeful will be
the case), the portion of parameter space
for which it is viable expands; \eg\ using $\etmin=50\gev$ at $L=30\fbi$,
the $\geq 5$ event region would extend to $\sqrt F\sim 1800-2300\tev$ 
for $\Lambda\lsim 120\tev$. 
\eit

\item The impact-parameter signal.

Requiring $\etmin=50,70\gev$ at $L=2,30\fbi$, respectively,
Figs.~\ref{contoursfullfigv} and \ref{contoursfullfigvii} show
that the $b$ signal has an even larger region with $\geq 5$ events than
does the OHC signal. For $L=2\fbi$, one finds 5 or more events
for $\sqrt F\lsim 1200-1700$ for $\Lambda\lsim 115\tev$. For $L=30\fbi$,
one finds 5 or more events for $\sqrt F\lsim 2200-3000\tev$ for $\Lambda\lsim
150\tev$.  Only in a small corner with small $\sqrt F$
and large $\Lambda$ does the $b$ signal fail, and this region is handily
covered by the prompt-$2\gam$ signal. 
As compared to the tri-lepton signal, the impact-parameter signal extends
to substantially higher $\Lambda$ values for $\sqrt F\lsim 1600\tev$ 
at $L=2\fbi$ and for $\sqrt F\lsim 2700\tev$ for $L=30\fbi$.

From comparing Figs.~\ref{contoursfullfigiii}, 
\ref{contoursfullfigv} and \ref{contoursfullfigvii} we can see
that weakening the $L=30\fbi$ missing energy cut to $\etmin=50\gev$
results in a slight increase in the $\geq 5$ event parameter space region.
Weakening the cut still further to $\etmin=30\gev$
appears (we cannot be precise since our scan
did not extend to high enough $\sqrt F$) to result
in a substantial increase to higher $\sqrt F$ of the $\geq 5$ event region.

\item The roof-array signal.

For $L=2\fbi$, the RA signal has $\geq 5$ events 
up to $\sqrt F \lsim 1500\tev$ for $\Lambda\lsim 40\tev$.
In all of this region, the tri-lepton
and impact-parameter signals, but not the tri-lepton-plus-prompt-photon 
signal, will also be viable,
For $L=30\fbi$, the RA signal has $\geq 5$ events for 
large $\sqrt F\lsim 2800-3000\tev$ when $\Lambda\lsim 120\tev$. 
This $\Lambda$ reach at high $\sqrt F$ is substantially greater
than achieved by the tri-lepton signal.
As $\sqrt F$ decreases, the $\Lambda$ extent of the
$\geq 5$ event region decreases.

From comparing Figs.~\ref{contoursfullfigiii}, 
\ref{contoursfullfigv} and \ref{contoursfullfigvii} we can see
that weakening the $L=30\fbi$ cut to $\etmin=50\gev$,
or, if possible without encountering backgrounds, $\etmin=30\tev$,
substantially increases
the high $\sqrt F$ reach of the $\geq 5$ event region for $\Lambda\lsim
120\tev$.~\footnote{Since our parameter scan was limited
to $\sqrt F\leq 2800$, we are unable to be quantitative in this
statement.}

\eit

One especially
important point regarding the delayed-decay signals discussed above is 
the following. If $\sqrt F$ is large, they are the only signals that will allow
one to ascertain that nature has chosen a GMSB model, even if supersymmetry
is detected in the jets-plus-missing-energy and/or tri-lepton modes.
In particular, we saw above that 
the tri-lepton-plus-prompt-photon and prompt-two-photon
signals will not be in evidence at large $\sqrt F$,
regardless of the magnitude of $\Lambda$.
Without the delayed-decay signals, a SUGRA  model with GMSB-like boundary
conditions for the soft-supersymmetry-breaking masses of the
gauginos and sfermions would then be indistinguishable
from a GMSB model with a large value for $\sqrt F$.

\begin{figure}[p]
\leavevmode
\epsfxsize=5.5in
\centerline{\epsffile{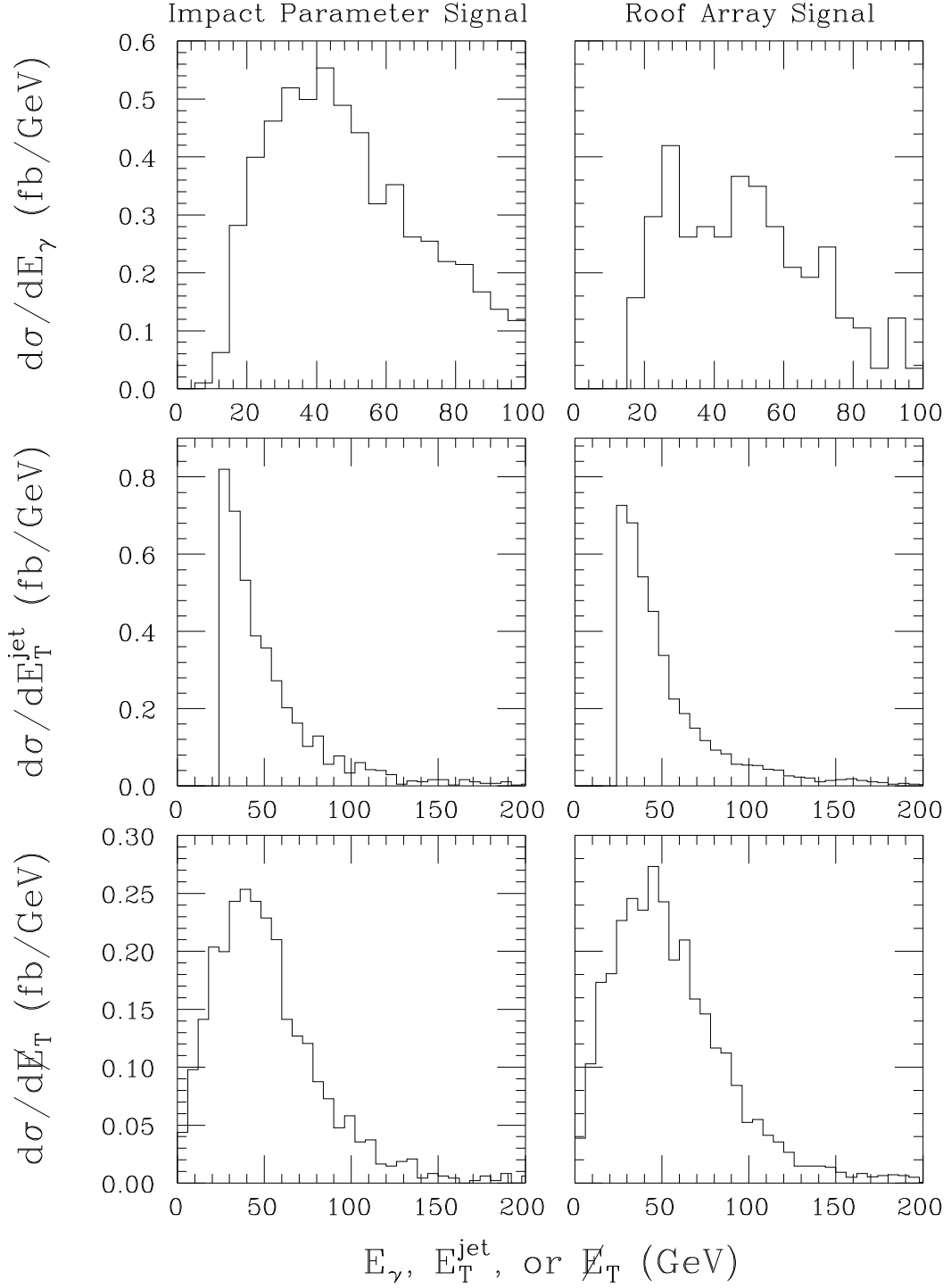}}
\bigskip
\caption{\baselineskip=0pt  
For the impact-parameter and roof-array signals,
we plot $d\sigma/dE_\gam$, $d\sigma/d\etjet$
and $d\sigma/d\etmiss$ for the parameter space point $\protect\sqrt F=1400\tev$
and $\Lambda=50\tev$. The limited cuts applied are described in the text.}
\label{eplotsfig}
\end{figure}

The dependence of our signals on various energy variables may be
useful in designing triggers and considering backgrounds.  To
this end, we present a figure showing the distributions in
$E_{\gam}$, $\etjet $ and $\etmiss$
for the impact-parameter and roof-array signals.
These distributions are presented for the parameter space point
$\sqrt F=1400\tev$ and $\Lambda=50\tev$; they are fairly representative
of the results obtained at other parameter space points.
In these distributions, we have imposed a more limited set of cuts.  
\bit
\item
In the case of the distributions presented
for the impact-parameter signal, we require only that there be one or
more delayed-decay impact-parameter photon (as defined earlier) with $b>2$ cm,
except that when plotting the $E_\gam$ distribution, we remove
the cut on $E_\gam$.
\item
In the case of the distributions presented for the roof-array signal,
we require only that there be one or more delayed-decay roof-array photon
(as defined earlier). 
\eit
In neither case do we impose
jet cuts, except that when plotting $d\sigma/d\etjet $,
the jets must satisfy the definition of a jet as given earlier.
In addition, no cuts are placed on $\etmiss$.
In the case of the $\etjet $ and $E_\gam$ distributions, all
jets and delayed-decay (impact-parameter in the first case, and
roof-array in the second case)
photons satisfying the minimum requirements noted above
are included in the histograms.

From the plots of Fig.~\ref{eplotsfig} we observe the following.
\bit
\item
Our $E_\gam$ cuts accept essentially all events and could be strengthened
somewhat without significant loss of signal event rate.
\item
The $\etjet $ spectrum is falling very rapidly, and it would
be undesirable to strengthen the jet cuts unless it is absolutely
necessary.
\item
Cuts significantly stronger than $\etmin>70\gev$ will start to
cause a significant loss of signal rate.  Thus, we hope that
such a stronger cut will not be needed to eliminate backgrounds.
\eit

\begin{figure}[h]
\leavevmode
\epsfxsize=5.5in
\centerline{\epsffile{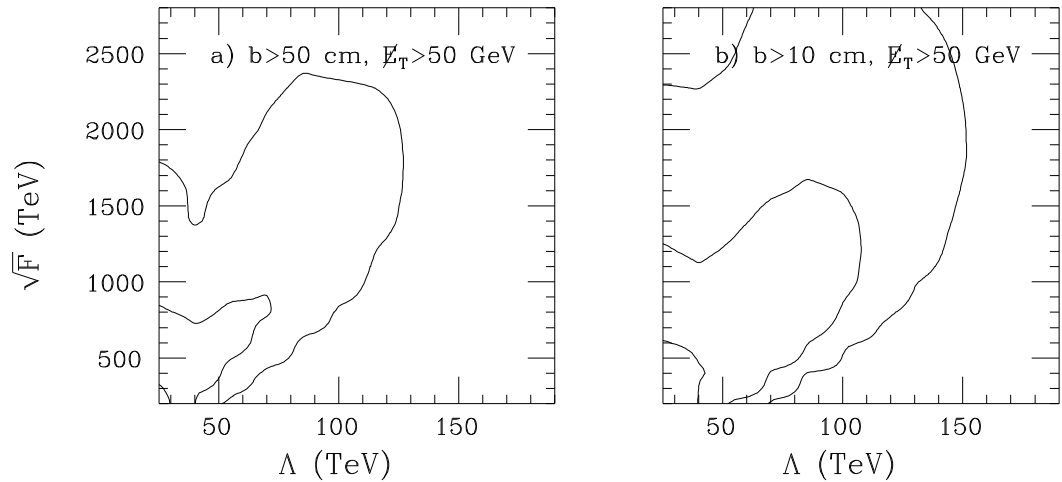}}
\bigskip
\caption{\baselineskip=0pt  
For $\etmiss>50\gev$, 
we plot the same cross section contours for the impact-parameter signal
as given in Fig.~\ref{contoursfullfigv} (for $b>2$cm), 
for the cases (a) $b>50$cm and (b) $b>10$cm.  All other cuts are identical.}
\label{bcutfig}
\end{figure}

It is also potentially important to know how sensitive the impact-parameter
signal is to the impact-parameter resolution.  To this end, we present in
Fig.~\ref{bcutfig} the same cross section contours for the impact-parameter
signal as given in Fig.~\ref{contoursfullfigv} (for $b>2$cm),
for the cases (a) $b>50$cm and (b) $b>10$cm, keeping
all other cuts (including $\etmin=50\gev$) the same. 
We observe that the $b$ cut can be increased from $2$cm to $10$cm
without significant damage to the signal. But, if
the transverse impact parameter resolution were as poor as 
the 5cm--10cm resolution applicable \cite{landsberg} during Run-I (for which
the D0 detector did not have a pre-shower device) a cut of $b>50$cm
would probably be needed to eliminate backgrounds and 
the region of viability of the impact-parameter signal would
decrease very substantially. In fact, for 
all cases [$(L=100\pbi,\etmin=30\gev)$,
$(L=2\fbi,\etmin=50\gev)$ and $(L=30\fbi,\etmin=70\gev)$]
the region for which $\geq 5$ events are found for the impact-parameter signal
after requiring $b>50$cm is quite similar to that 
for which $\geq 5$ OHC signal events are found. In contrast, the 
region for which $\geq5$ 
impact-parameter signal events are expected
for the $b>10$cm or $b>2$cm cuts is always substantially larger than
the region for which $\geq 5$ OHC signal events 
are predicted in all the above three $(L,\etmin)$ cases.

\section{Discussion and Conclusions}

We have studied a simple gauge-mediated supersymmetry breaking model
in which the lightest neutralino, the $\cnone$,
is the next-to-lightest supersymmetric particle
and decays to photon plus gravitino. The model
is characterized by the parameters $\Lambda$ (which
sets the scale of the masses of the superpartners of the Standard
Model particles) and $\sqrt F$ (which specifies the scale at
which supersymmetry is broken in the hidden sector and, thereby, determines
the mass of the gravitino and the lifetime of the $\cnone$).
We have argued that it is very possible (indeed,
preferred in the simplest existing GMSB models)
for the $\sqrt F$ parameter to be 
sufficiently big that there is a large probability for
the $\cnone$ to decay a substantial distance from the interaction point.
We have shown that the portion of the $(\sqrt F,\Lambda)$ 
parameter space for which
a signal for supersymmetry can be seen at the Tevatron
is greatly expanded by employing signals sensitive to such delayed decays.

In particular, it is useful to compare (a) the standard
SUSY signatures (jets plus missing-energy or three leptons plus
missing-energy), (b) the prompt-two-photon signal and (c) the delayed-decay
(impact-parameter, roof-array and 
outer-hadronic-calorimeter) signals. 
\begin{itemize}
\item
Of the standard SUSY signatures, the tri-lepton signal is the stronger.
Using it, discovery of GMSB supersymmetry will be possible 
for small to moderate $\Lambda$ values, $\lsim 65$ ($\lsim 75\tev$) 
at Run-II (Tev33), regardless of the size of $\sqrt F$.
\item
The prompt-two-photon signal is viable for small to moderate
$\sqrt F$, but extending to reasonably large $\Lambda$,
$\Lambda\lsim 100\tev$ ($\lsim 150\tev$) at Run-II (Tev33). 
\item 
The delayed-decay signals 
are potentially viable in the large-$\sqrt F$,
large-$\Lambda$ portion of parameter space not covered
by the standard SUSY and prompt-two-photon signals. 
For Run-II, the delayed decay signals could
allow supersymmetry discovery for all of $\sqrt F\lsim 1700\tev$,
$\Lambda\lsim 100\tev$, extending at Tev33
to cover essentially the entire $\sqrt F\lsim 3000\tev$
and $\Lambda\lsim 120\tev$ region that is preferred in the context
of the model we have explored.~\footnote{In a more general GMSB model,
it is certainly possible that the entire interesting region
of parameter space might not be accessible using the delayed-decay
signals. Still, we expect that the delayed-decay
signals will always significantly expand the amount of GMSB parameter
space at larger $\sqrt F$ for which a supersymmetry signal can be discovered.}
\end{itemize}

If $\sqrt F$ is large, the delayed-decay signals will be important
even if $\Lambda$ is in the small to moderate range ($\Lambda\lsim
65-75\tev$) for which
the standard jets-plus-missing-energy and/or tri-lepton
SUSY signatures will be visible.
This is because the prompt-photon signals (tri-lepton-plus-prompt-photon
or prompt-two-photon) will not be in evidence
if $\sqrt F$ is big.  Without employing
the impact-parameter, roof-array and/or outer-hadronic-calorimeter
delayed-decay signals there will be no hint
that nature has chosen a GMSB model, as opposed to a SUGRA model
with GMSB-like boundary conditions for gaugino and sfermion 
soft-supersymmetry-breaking masses.

If delayed decay signatures are seen, the next goal will be to determine
the parameters of the GMSB model.  These include the all important
value of $\sqrt F$ and the masses
of the supersymmetric particles. The latter will allow
us to determine the overall scale $\Lambda$ and
the relative sizes of the soft-supersymmetry-breaking masses
as a function of $\Lambda$.  Important input information will include:
(a) the distribution of $\cnone$
decays as a function of distance from the interaction point,
as reflected in the relative rates for different signals (\eg\
the prompt-two-photon rate as compared to the roof-array rate);
(b) the energy distribution of the photons from the $\cnone$ decays;
(c) the time at which the photons from the $\cnone$ decays arrive
at various elements of the detector (\eg\ the electromagnetic
calorimeter pre-shower, the outer-hadronic-calorimeter cell or the roof-array);
(d) the missing-energy distribution; and
(e) the types of different particles that appear in association with
the supersymmetric events, and their energy distributions.
All of these depend in a complicated way on the parameters of interest.
For example, the distance distribution of $\cnone$ decays, the photon energies,
and the timing of the photons from the decays all
depend strongly on $(c\tau)_{\cnone}$ and $\mcnone$,
and might allow their determination.~\footnote{As an example, one
could measure the time of flight of the $\cnone$ to the OHC cell 
(by the time of the OHC energy deposit) and 
the energy of the deposit (\ie\ the energy of the photon).
The flight time is a function of $c\tau$, $\mcnone$ and the average $\cnone$
energy. The photon energy is a function of the latter two. 
The correlation between the flight time and photon energy (as measured in
a large number of events) might allow
a roughly model-independent determination of $c\tau$ and $\mcnone$.}
Even a rough measurement of these latter
two quantities will allow us to estimate
$\sqrt F$ using Eq.~(\ref{ctauform}).
Further,
the energy distributions of the visible particles (jets, leptons, and photons)
will reflect both the relative and absolute masses
of the sparticles. For example, 
thresholds reflecting $\mslepr-\mcnone$ and/or $\mslepl-\mcnone$
might show up in lepton energy distributions. And so forth. Obviously,
substantial statistics will be required for a relatively model-independent
determination of $\sqrt F$ and the soft-supersymmetry-breaking masses.

We note that all the delayed-decay signals we considered here
are potentially useful for
any model in which the decay of the $\cnone$ is not prompt.
For example, in R-parity violating (RPV) SUGRA models the RPV couplings
may be sufficiently small that the $c\tau$ for $\cnone$ decay
is quite substantial.  The products emerging from the $\cnone$ decay
(whether three leptons, one lepton plus two jets, or three jets)
can emerge in the outer hadronic calorimeter, with large impact parameter,
or outside the detector and pass through the roof array.
There is also the possibility that R-parity violation could be present
in GMSB models.  For an appropriate balance between
the magnitude of $\sqrt F$ and the size
of the RPV couplings, the $\cnone$ could have prominent decays
both to RPV channels and to $\gam\gtino$, and both types of $\cnone$
decays could be substantially delayed. In general, observation
of the delayed-decay
signals would be quite critical to determining the relative branching
ratios for RPV and $\gam\gtino$ decays of the $\cnone$. A similar set
of remarks would apply to RPV (SUGRA or GMSB) models in which a right-handed
slepton is the lightest of the superpartners of the SM particles.~\footnote{The
main difference, as compared to the $\cnone$ case, is that
additional information would arise from observation of
the visible track associated with a semi-stable charged $\slepr$. Also,
a long-lived $\slepr$ might be absorbed inside the detector.}
Thus, the utility of the types of delayed-decay
signals we propose will extend far beyond the NLSP=$\cnone$,
R-parity-conserving GMSB model context analyzed in detail here.

Our study has focused on using the D0 detector in the configuration
planned for Run-II, with the addition of a (relatively cheap) roof-array
detector.  There are, of course, much more ambitious possibilities in which
a decay volume and special detector, specifically designed 
to pick up delayed $\cnone$ decays, is constructed some
distance from the main detector. Nonetheless, since the
delayed-decay photon signals discussed here show great promise 
and since delayed decays in general could easily prove to be the hallmark
signature of gauge-mediated supersymmetry breaking, not to mention
R-parity violation,
the CDF and D0 detector groups should work to clarify and refine
these signals and, in particular, the backgrounds thereto.

\vspace{1.5cm}
\centerline{\bf Acknowledgments}
\vspace{.5cm}
This work was supported in part by the U.S. Department of Energy under
grant No. DE-FG03-91ER40674, and
by the Davis Institute for High Energy Physics. We are grateful 
to B. Dobrescu, H. Murayama, and S. Thomas for helpful conversations.
We particularly wish to acknowledge important conversations
with S. Mani regarding the D0 detector and the possibility of
implementing the delayed-decay signals discussed here.


\end{document}